\documentclass[
    ,final            
  ]
  {aipproc}

\layoutstyle{6x9}

\newcommand{\la}{\raisebox{-0.6ex}{$\,\stackrel
{\raisebox{-.2ex}{$\textstyle <$}}{\sim}\,$}}
\newcommand{\ga}{\raisebox{-0.6ex}{$\,\stackrel
{\raisebox{-.2ex}{$\textstyle >$}}{\sim}\,$}}

\begin{document}

\title{Understanding selection effects in observed samples of cataclysmic variables}


\author{Magaretha L. Pretorius}{
  address={School of Physics and Astronomy, University of Southampton, Hampshire SO17 1BJ, UK}
}

\author{Christian Knigge}{
  address={School of Physics and Astronomy, University of Southampton, Hampshire SO17 1BJ, UK}
}

\author{Ulrich Kolb}{
  address={Department of Physics and Astronomy, The Open University, Milton Keynes MK7 6AA, UK}
}

\begin{abstract}
Large differences between the properties of the known sample of cataclysmic variable stars (CVs) and the predictions of the theory of binary star evolution have long been recognised.  However, because all existing CV samples suffer from strong selection effects, observational bias must be considered before it is possible to tell whether there is an inconsistency, which would imply a failure of the evolutionary model.  We have modelled common selection effects and illustrate their influence on observed CV samples.
\end{abstract}

\maketitle


\section{Introduction}
Cataclysmic variables (CVs) are semi-detached binary stars with orbital periods ($P_{orb}$) typically of the order of hours, consisting of a white dwarf primary accreting from a companion which is usually a late-type, approximately main-sequence star.  See \cite{bible} for a comprehensive review of the subject.

Binary star evolution is driven by loss of orbital angular momentum through gravitational quadrupole radiation, and, in at least some cases, the much more efficient mechanism of magnetic braking \cite{perm1,p17,vz,perm3}.  Observational estimates of the mass transfer rate ($\dot{M}$) are on average much higher for long-period systems ($P_{orb}\ga 3$~h) than for systems at $P_{orb}\la 3$~h \cite{pat1}, suggesting that mass transfer is driven mainly by magnetic braking in long-period systems, while gravitational radiation dominates at $P_{orb}\la 3$~h.

\begin{figure}
 \includegraphics[width=84mm]{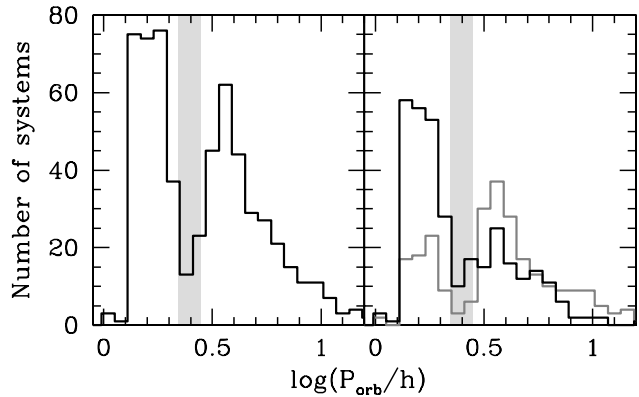}
 \caption{The orbital period distribution of non-magnetic, hydrogen-rich CVs (left hand panel) \cite{rkcat}.  Right hand panel: the same distribution divided into two magnitude bins, $V > 16.5$ (black) and $V \le 16.5$ (dark grey).  The period gap is indicated by shading of the range $2.2\,\mathrm{h} < P_{orb} < 2.8\,\mathrm{h}$ in both panels.
}
 \label{fig:porb}
\end{figure}

The orbital period distribution of CVs is a useful indicator in the study of their evolution because loss of angular momentum, and the resulting mass transfer, changes $P_{orb}$.  This distribution is shown in the left hand panel of Fig.~\ref{fig:porb}. Its most striking features are the period gap and period minimum: a pronounced drop in the number of systems at $2.2~\mathrm{h} \la P_{orb}\la2.8$~h and a sharp cut-off at about 80~min.  The period gap is usually explained as resulting from disrupted magnetic braking \cite{perg1,perg2,perg3}.  The period minimum is a consequence of the response of a secondary of very small mass to continuing mass loss \cite{perm1,p17,perm3}.  When the thermal time-scale of the secondary exceeds the mass-transfer time-scale, the star is not able to shrink fast enough (or in fact expands) in response to mass loss, so that the orbital evolution moves back through longer periods.  CVs in this phase of evolution are referred to as `period bouncers'. 

The rudiments of the theory of CV evolution have long been accepted, yet there are still significant discrepancies between the predictions of theory and the properties of the observed CV sample.  Population synthesis studies predict that $\simeq 99$\% of the intrinsic galactic CV population should be short-period systems, and that about 70\% should be period bouncers \cite{pop,howell2}.  A glance at Fig.~\ref{fig:porb} is enough to show that the first of these predictions is  not reflected in the known sample of CVs.  Furthermore, only a handful of known systems are likely period bouncers \cite{pat4}.  Clearly, if the standard theory of CV evolution is even approximately correct, the observed CV sample is not representative of the intrinsic population.  However, since different types of CVs differ in intrinsic brightness and all surveys are flux limited, observed samples are not expected to reflect the intrinsic population.  Apparent brightness and large amplitude variability are two factors that the discovery probability obviously depends on; since both intrinsic brightness and the frequency of large amplitude brightness variations decrease with $\dot{M}$, short-period, low-$\dot{M}$ systems are necessarily under-represented in the known sample.  The question is only whether these and other selection effects can account for the size of the discrepancy between the period distributions of the predicted and observed samples.  The right-hand panel of Fig.~\ref{fig:porb} goes some way towards illustrating the importance of the apparent magnitudes of known CVs.

We use a Monte Carlo approach to study the influence of simple but important selection effects on observational CV samples.

\section{The computational method}
A population of CVs drawn from the distribution in $P_{orb}$, $\langle \dot{M} \rangle$, $M_1$, and $M_2$ (where $\langle \dot{M} \rangle$ is the long-term average mass transfer rate, and $M_1$ and $M_2$ are the primary and secondary mass) resulting from model pm5 of \cite{pop} is distributed in a model galactic disc with exponential density profiles.  The vertical scale heights are 120~pc for long-$P_{orb}$ systems, 260~pc for normal short-$P_{orb}$ systems, and 450~pc for period bouncers \cite{howell,rc86}.  We take the galactic centre distance as $7\,620\,\mathrm{pc}$ \cite{gcd}, and compute interstellar extinction by integrating the density of the interstellar medium \cite{ism} along each line of sight.

In calculating the spectral energy distribution of each system, we account for the accretion disc, bright spot, white dwarf, and secondary star.  The bright spot is modelled as a blackbody.  The flux contributions of the secondary and primary are found using the spectral type -$P_{orb}$ relation of \cite{sd} and the $T_{eff}$--$\langle \dot{M} \rangle$ relation of \cite{tb}.  We used the code of \cite{disc} to compute a 4D grid of accretion disc models over the variables $\dot{M}_d$ (the rate at which mass flows through the disc), $M_1$, the orbital inclination $i$, and the outer disc radius $r_d$.  Most CVs have discs that are subject to a thermal instability; these systems are the dwarf novae (DN).  The mass accretion rate through the disc in outburst ($\dot{M}_{dO}$) and quiescence ($\dot{M}_{dQ}$) are related by $\langle \dot{M} \rangle = C \dot{M}_{dO} + (1-C)\dot{M}_{dQ}$, where $C$ is the outburst duty cycle.  Empirical estimates of the absolute $V$ magnitude at maximum and a bolometric correction of $-1.8$ yield the disc luminosity ($L_d$); we set $\dot{M}_{dO}=2R_1L_d/GM_1$.  Then $\dot{M}_{dQ}$ follows from assuming $C$, and the probability of catching a DN in outburst in a single-epoch survey equals $C$.  $r_d$ is set to $0.7 R_L$ for nova-like variables and DN in outburst, and $0.5 R_L$ for quiescent systems \cite{haw}, where $R_L$ is the Roche-lobe radius of the primary.

\section{Selection effects}
The single characteristic through which the most CVs have been discovered is large amplitude variability (DN outbursts and nova eruptions).  Most other CVs were found as objects with blue optical colours or X-ray emission, mainly in severely flux-limited surveys.  A quantitative comparison between theory and observations requires a reasonably large observed sample with well defined selection criteria.  We limit ourselves here to a mostly qualitative investigation of a few important selection effects.

\subsection{Optical flux limits}
\begin{figure}
 \includegraphics[width=148mm]{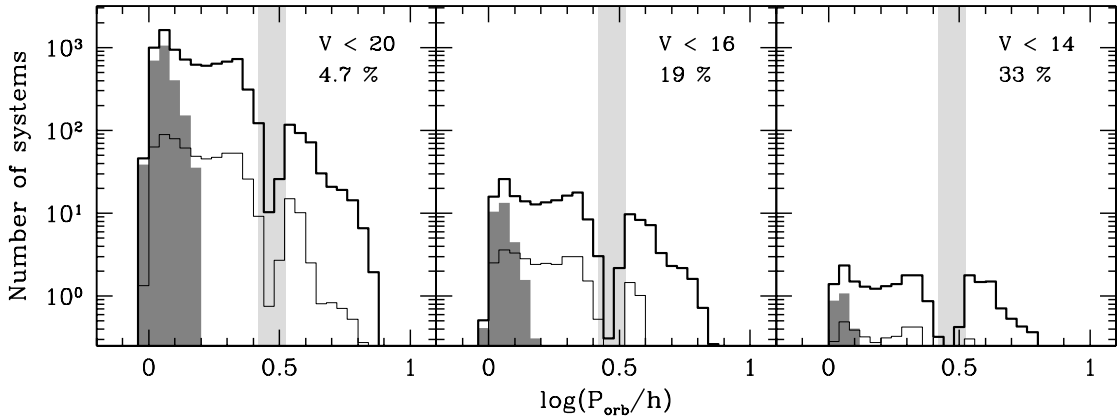}
 \caption{Period distributions of magnitude limited samples.  The magnitude limit is given in each panel, together with the fraction of the sample with periods above the period gap.  Shading of the range $2.63\,\mathrm{h} < P_{orb} < 3.35\,\mathrm{h}$ indicates the model period gap.  The contribution of period bouncers is shaded in dark grey; the fine black histograms show DN that were simulated in outburst.  
}
 \label{fig:flim}
\end{figure}
\noindent
Bias towards finding apparently bright objects is the most simple and important selection effect (see e.g. \cite{rit1}).  Fig.~\ref{fig:flim} displays period histograms of samples with three different magnitude limits ($V<20$, $V<16$, and $V<14$).  In this figure (and all other period histograms we will show) the number of systems in each bin is scaled to reproduce a local space density of $5 \times 10^{-5}\,\mathrm{pc^{-3}}$.  The shape of the period distribution depends strongly on the magnitude limit.  As is expected from the fact that long-period CVs are intrinsically brighter, the fraction of the total sample made up by long-period systems increases at brighter magnitude limits; it is 4.7\%, 19\%, and 33\% for $V<20$, 16, and 14 respectively.  

Of course the known CV sample is not complete to 20th or even 16th magnitude.  In fact, comparing the magnitude distribution of known CVs to a complete model sample indicates that the known sample is probably incomplete even at $V=13$.  Therefore the distributions in Fig.~1 should not be compared directly to the known sample.

\subsection{Blue optical colours}
\begin{figure}
 $\begin{array}{c@{\hspace{0.8cm}}c}
 \includegraphics[width=70mm]{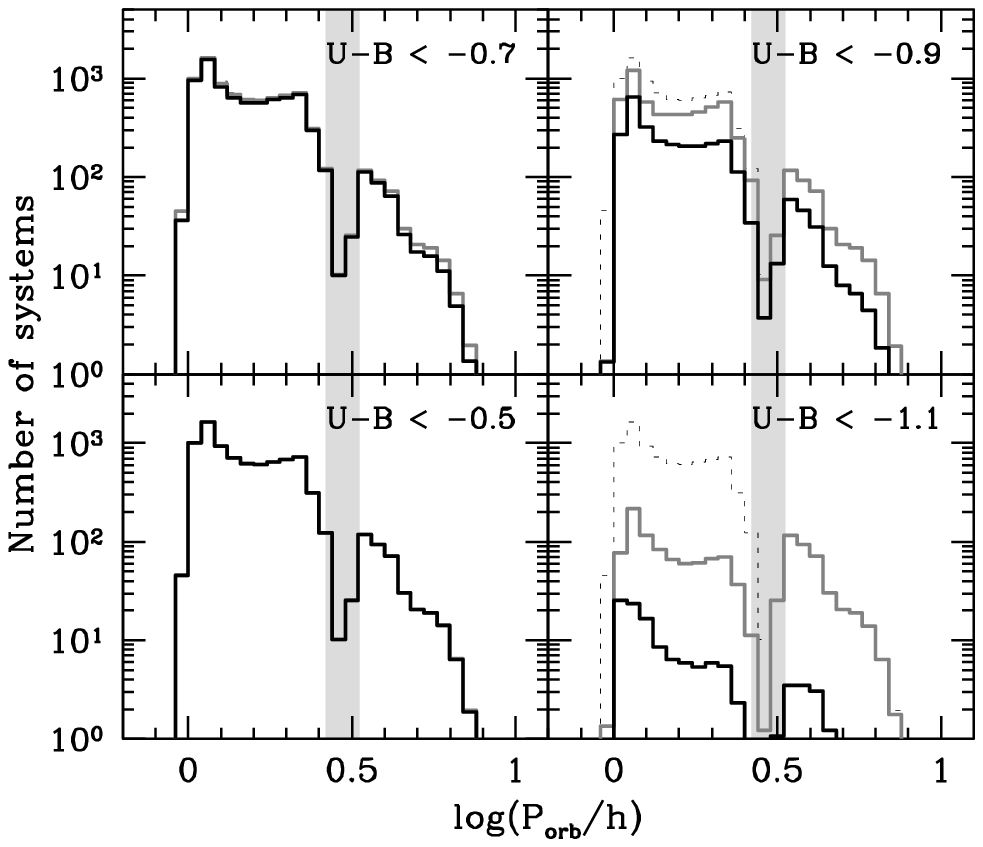} &
 \includegraphics[width=70mm]{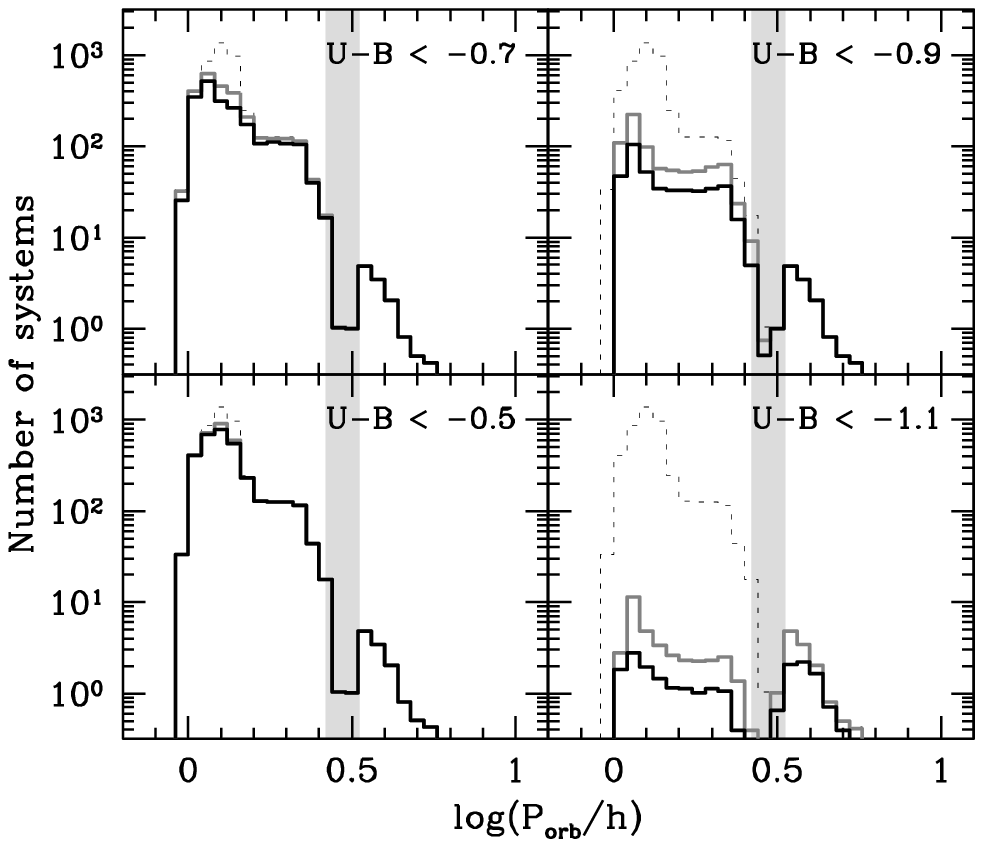}  \\ [0.1cm]
\parbox{70mm}{{\footnotesize \textbf{FIGURE~3.} Period histograms of systems with $V<20$ (dotted), and systems with $V<20$ and the $U-B$ indicated in each panel (black).  Dark grey histograms show the samples obtained when the same blue cuts are applied to the intrinsic colours.  Almost all systems with $V<20$ also have both $(U-B)_0<-0.7$ and $U-B<-0.7$, and most long-period CVs have $(U-B)_0<-1.1$.}
} &
\parbox{70mm}{{\footnotesize \textbf{FIGURE~4.} Period histograms of samples with the same $U-B$ and $(U-B)_0$ cuts as in Fig.~3, but here applied to a volume limited sample.  Dotted histograms show the complete sample; black are samples with observed colours satisfying the blue cut indicated in each panel, and grey shows systems with intrinsic colours blue enough to satisfy the selection cuts.}
}
 \end{array}$
\end{figure}
\noindent
Most known CVs have $(U-B)_0<-0.5$ \cite{colours}.  This is because high-$\dot{M}$ systems have high continuum colour temperatures, while low-$\dot{M}$ systems have the Balmer discontinuity in emission.  However, in the systems with lowest $\dot{M}$, the cool white dwarf is expected to dominate the optical flux.  Blue selection cuts may therefore exclude the faintest CVs.

Fig.~3 shows the effect of a magnitude limit together with different selection criteria in $U-B$.  In every panel the dotted histogram shows all systems with $V<20$, and black shows systems that also satisfy the blue cut.  Only at $U-B<-0.9$ does the blue selection start removing a large fraction of systems from the magnitude limited sample.  Fig.~4 shows $P_{orb}$ distributions resulting from blue cuts imposed on a volume limited sample (produced in such a way that all CVs have the same spacial distribution, so that this sample is representative of the intrinsic population).  It is seen that even $U-B<-0.5$ excludes systems near the period minimum; $U-B<-0.7$ introduces a severe bias against short-period systems.  In the samples shown in Fig.~3, most of these systems were already excluded by the magnitude limit.  A survey that selects objects for $U-B \la -0.7$ is thus expected to be seriously biased against short-period, low-$\dot{M}$ CVs, but the colour cut introduces hardly any additional bias in a survey that is also severely flux limited, as is the case for existing $UV$-excess surveys.

The grey histograms in Fig.~3 and 4 display samples with the same blue cuts as shown by black histograms, but here the cuts are applied to the intrinsic colours to illustrate the effect of reddening.  Notice that practically all long-period systems have $(U-B)_0<-0.9$, but some get excluded from observed samples because they are on average more distant than short-period systems, and thus reddened.  This is amplified by the fact that they are also more concentrated towards the galactic plane.

\subsection{Restrictions in galactic latitude}

\setcounter{figure}{4}
\begin{figure}
 \includegraphics[width=84mm]{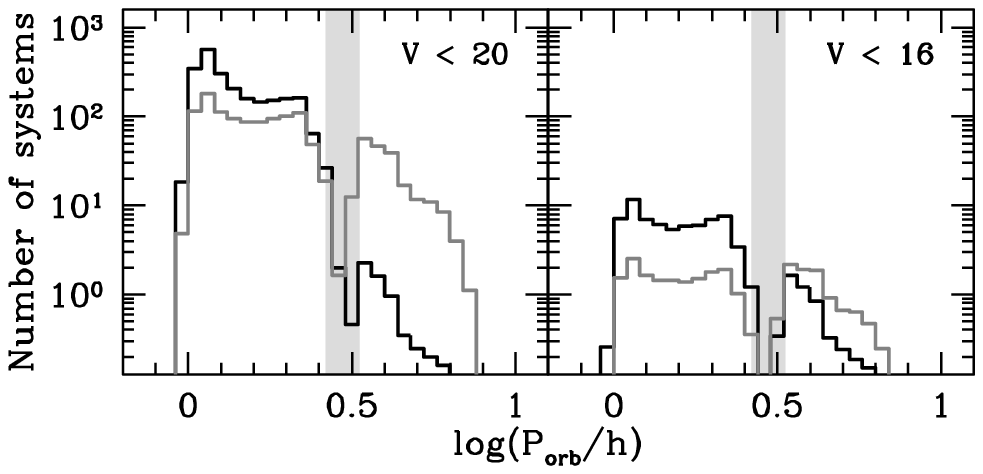}
 \caption{The orbital period distributions of high galactic latitude ($|b|>30^\circ$; black) and galactic plane ($|b|<5^\circ$; dark grey) samples with magnitude limits $V<20$ (left) and $V<16$ (right).  }
 \label{fig:gb}
\end{figure}

\noindent
Surveys at high galactic latitude hold some promise of uncovering samples that are volume limited.  In a galactic plane survey, on the other hand, intrinsically bright objects at large distances will be included in disproportion to their space density.  The direction of this effect is obvious; we simply show its magnitude in Fig.~\ref{fig:gb}, with period histograms of high (black) and low (dark grey) galactic latitude samples with two magnitude limits.  As expected, the ratio of short- to long-period systems is smaller in the galactic plane samples, and this effect is larger for the deeper samples.

In a survey covering the area $|b|>20^\circ$, less than 10.0\% of CVs in the survey volume are detected above a magnitude limit of $B<20$.  This implies that even the CV sample produced by a deep, high-$b$ survey such as the Sloan Digital Sky Survey \cite{sdsscvs} will by no means be volume limited.  Note also that, since the vertical galactic distribution of CVs differ for sub-populations with different typical ages, a volume limited, high-$b$ sample is in any case not the same as the intrinsic CV population.

\section{Conclusions}
We have presented a Monte-Carlo technique to model selection effects in observed samples of CVs.  Given a theoretical intrinsic population and any survey with well defined selection criteria, this technique can be applied to predict the observed population.  

For practically achievable flux limits, no flux limited sample is representative of the intrinsic CV population; it is therefore quite inappropriate to compare an observed sample with the predicted intrinsic population.  The properties of the period distribution of a magnitude limited sample depend strongly on the magnitude limit.  A selection cut of $U-B\la-0.7$ introduces a serious bias against detecting short-period systems, but this is unimportant in the presence of a bright flux limit.  Finally, magnitude limited surveys at high galactic latitude are expected to produce samples very different from galactic plane surveys, because they detect a larger fraction of all systems inside the volume defined by the galactic latitude range.


\begin{theacknowledgments}
MLP acknowledges financial support from the South African National Research Foundation and the University of Southampton.  We thank Romuald Tylenda for the use of his accretion disc model.
\end{theacknowledgments}

\bibliographystyle{aipproc}   

\end{document}